# The binary asteroid 22 Kalliope: Linus orbit determination on the basis of speckle interferometric observations

I.A. Sokova[1*], E.N. Sokov[1], E.A. Roschina[1], D.A. Rastegaev[2], A.A. Kiselev[1], Yu. Yu. Balega[2], D.L. Gorshanov[1], E.V. Malogolovets[2], V.V. Dyachenko[2], A.F. Maksimov[2]

[1] *The Pulkovo Astronomical Observatory (CAO RAS),*
*Pulkovskoye chaussee, 65, Saint-Petersburg, Russia 196140*

[2] *The Special Astrophysical Observatory (SAO RAS),*
*Special Astrophysical Observatory, Nizhnij Arkhyz, Zelenchukskiy region,*
*Karachai-Cherkessian Republic, Russia 369167*

**Abstract**

In this paper we present the orbital elements of Linus satellite of 22 Kalliope asteroid. Orbital element determination is based on the speckle interferometry data obtained with the 6-meter BTA telescope operated by SAO RAS. We processed 9 accurate positions of Linus orbiting around the main component of 22 Kalliope between 10 and 16 December, 2011. In order to determine the orbital elements of the Linus we have applied the direct geometric method. The formal errors are about 5 mas. This accuracy makes it possible to study the variations of the Linus orbital elements influenced by different perturbations over the course of time. Estimates of six classical orbital elements, such as the semi-major axis of the Linus orbit $a = 1109 \pm 6$ km, eccentricity $e = 0.016 \pm 0.004$, inclination $i = 101° \pm 1°$ to the ecliptic plane and others, are presented in this work.

## Introduction

The existence of binary asteroids in the Solar system was merely a hypothesis just at the end of the twentieth century. Today there is no doubt that binary asteroids exist thanks to the revolutionary discovery of the binary asteroid 243 Ida by the space mission Galileo at August 28, 1993 year. Scientists have discovered not only binary, but also triple asteroid systems in the main asteroid belt as well as among near-Earth asteroids (the NEA group) and even in the Kuiper belt. The discovery of binary and multiple asteroid systems has allowed us to consider a different perspective on the origin and formation of solar system objects.

Understanding the origin of binary and multiple asteroid systems and their dynamics is still a serious problem. Research in this area helps to shed light on such global issues as solar system formation and evolution. Thus, the existence of binary and multiple asteroid systems can attest to the fact that at the early stage of the solar system formation the asteroid belt contained larger "parent" bodies which later disintegrated into smaller ones. The study of binary asteroids from the NEA group allow us to assume that the objects from this group might be "descendants" from the main asteroid belt [Morbidelli et al, 2006]. However, binary and multiple asteroid systems have also been discovered in the Kuiper belt. There are lot of open questions in the area of the origin of small planetary bodies in our solar system.

---

*Corresponding author, e-mail: Iraida.Anna@gmail.com (Iraida Sokova)
Tel. +7(812)363 72 10, +7(960) 239 96 94

The present work is dedicated to the study of a binary asteroid 22 Kalliope located in the main asteroid belt. This is a typical binary asteroid from the main belt, but at the same time its size and configuration relative to the Earth have been allowed us to study it by applying direct imaging methods that are impossible to do with many other binary asteroids. Hence, the difference in brightness between the main component 22 Kalliope and its satellite Linus amounts to $3^m$, and the average angular distance between the components makes up approximately $0".3$. This asteroid is being actively studied by many scientists [Marchis et al, 2003; Descamps et al, 2008; Vachier et al, 2012 and others]. In particular, according to the preliminary determination of the Linus orbit researchers [Descamps et al, 2008] predicted mutual phenomena (eclipses and occultations) in the 22 Kalliope and Linus binary asteroid system. The photometric observations of these phenomena have made it possible to determine the shapes and sizes of the components as well as their densities. The main component 22 Kalliope was found to have a size of 181 km [Tedesco et al, 2002], but upon review and based on the observed eclipses and occultations the researchers have determined the size that makes up $166.2 \pm 2.8$ km, and the size of its satellite amounts to $28 \pm 2$ km [Descamps et al, 2008]. These estimates were confirmed by the NEOWISE observations ($D = 167.0 \pm 15.3$ km) [Masiero et al, 2011].

A recent paper on the determination of Linus' orbit based on the 10-year-long observation of the 22 Kalliope asteroid with large telescopes equipped with adaptive optics has been presented in [Vachier et al, 2012]. The authors have obtained the mean values of the Linus orbital elements for the last 10 years and concluded that the satellite orbit is stable within the accuracy limit of 15 mas.

In this work we determine the orbit of Linus based on speckle interferometric observations. Authors of this paper have obtained images of 9 positions of Linus based on the observed data collected at the 6-m BTA telescope operated by SAO RAS with the speckle interferometry method. The observations were performed during 5.2 days, i.e. it took less than two consecutive Linus revolutions around 22 Kalliope.

In order to determine the satellite orbital elements, we used the direct geometric method [Kisselev A. A., 1990; Kisselev A.A., 1997]. This method was developed for computing of a true orbit ellipse by using its apparent ellipse. The direct geometric method has been successfully applied for orbital determination of binary stellar systems [Grosheva E., 2006].

Similar to this method of determining orbit parameters in binary systems, is the applied geometric method developed by Pascal Descamps [Marchis et al., 2003; Descamps, P., 2005]. The main feature of our approach is an instantaneous determination of the satellite orbit. Recent research of the 22 Kalliope binary system [Vachier et al, 2012] has determined Linus' orbit using observations from the past 10 years, while in this research we have determined the Linus orbit based on observations performed over a few days. We do not eliminate Linus orbital variations as its orbit can be perturbed as a result of the main component 22 Kalliope's asymmetric and non-spherical shape, and as a result of periodic and non-periodic gravitational perturbations from large solar system bodies. Such perturbations have not been studied before, but a set of the methods presented in this work will probably allow for additional future research.

## Methods of multiple asteroid systems observations

Despite of the progress in the field of the solar system bodies research by the means of space probes and space telescopes, ground-based observations are still vital [Cellino, 2005]. Nowadays, a photometric method is commonly applied to the research of asteroid systems including binary and multiple asteroids as it allows us to determine the shape as well as the reflective properties of the asteroid's surface [Kaasalainen et al, 2002]. Spectral observations are also significant for determination of asteroids chemical composition and radar observations that are used to determine shapes and rotation periods of asteroids.

The methods used for obtaining object images, provide relative positions for binary asteroids. The direct imaging method with the use of adaptive optics is currently one of the most widespread methods for investigation of the large asteroids from main belt including binary and multiple asteroid systems. This method helped discover and confirm the main belt asteroids in binary and multiple systems, and moreover reliably determine orbits of asteroids' satellites for the first time [Merlin et al, 2000; Marchis et al, 2004; Marchis et al, 2005; Marchis et al, 2006; Marchis et al, 2008]. For example, mutual phenomena in the 22 Kalliope – Linus system in 2007 and 2012 have been predicted on the basis of orbit determination from direct imaging by observations with adaptive optics. [Descamps et al, 2008; Vachier et al, 2012]. Based on the mutual phenomena photometric observations carried out within the specified time period, the measurement accuracy of parameters of this binary system, such as the masses and sizes of both components has been improved.

An alternative method that allows obtaining of asteroid images is the speckle interferometry method. It is based on the processing of a large number of short-exposure images that allows for the "elimination" of possible atmospheric effects. This method is mostly applied the binary and multiple star systems observation as it makes it possible to obtain separate images of their components. The possibility of this method application to the binary and multiple asteroid systems investigations had been expressed by various researchers as far back as 20 years ago [Prokof'eva et al, 1995].

The speckle interferometry method has been applied to the observations of some binary and multiple asteroids and asteroids that were suspected of being binary or multiple in following works [Roberts et al, 1995; Cellino et al, 2003]. Obtained observations allowed to determine the shapes and the sizes of the asteroids. But there was no possibility of obtaining the images of the satellites of the asteroids from the above mentioned studies.

In this work we present the results of the binary asteroid satellite observation for the first time obtained by the means of the speckle interferometry method.

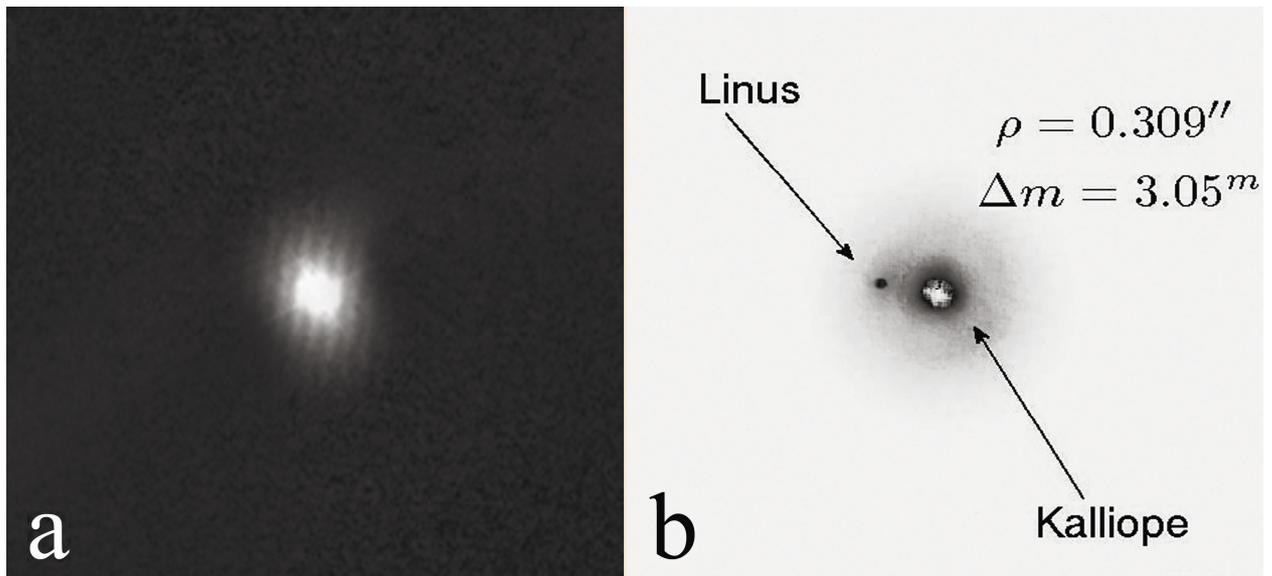

Fig. 1. a) The example of power spectrum of the binary asteroid 22 Kalliope obtained with the use speckle interferometer at the 6-m BTA telescope. b) The reconstructed image of satellite Linus and its main component 22 Kalliope. Angular distance between the components and apparent magnitude differences are given.

**The speckle interferometry method in detail**

The main contribution to the optical images distortion and blurring belongs to the atmospheric turbulence (or atmospheric seeing). For example, for a 6-m aperture of the optical BTA telescope at the wavelength of 550 nm the diffraction limit of a point source resolution has to equal 0″.02, while the real size of the image influenced by the atmospheric effects makes up 1-2″, i.e. 100 times more. The speckle interferometry method is a method of astronomical objects observation seen through turbulent atmosphere with the angular resolution limit close to the diffraction limit.

The principle of the speckle interferometry method is to take high-resolution images with a very short exposure time ($\sim 10^{-2}$ s). Such images consist of a great number of speckles that are produced by the mutual interference of the light beams that fall on the focal plane of a telescope from the different parts of the lens. Each speckle looks like an airy disk in the focal plane of a perfect telescope that is not affected by the atmospheric seeing. Atmospheric seeing influences the image in such a way that a wavefront that reaches a ground-based telescope is always distorted by the optical imperfections of the atmosphere.

When taking very short-exposure images we record the speckles distribution at that very instant, while with long exposure the image loses its structure and becomes blurred. In the images of a non-point (extended) source, the speckle pattern (their shape and size) reflects the characteristics of the source itself. For example, if we observe a binary object (a binary star or a binary asteroid), then the speckles are located in pairs, and each pair of speckles represents airy disk from the two components of a binary star or asteroid. In order to obtain information about the structure of the observed object we accumulated thousands of its snapshots. The special algorithms are used to obtain the relative positions in binary systems [Balega et al, 2002, Maksimov et al, 2003; Balega et al, 2011]. The power spectrum is built on the base of accumulated speckle interferograms and, then, is approximated by sinusoidal two-dimensional model using the least squares method [Pluzhnik, 2005]. The power spectrum for 22 Kalliope is shown on Fig. 1a. The relative coordinates of Linus are calculated from the parameters of the obtained model. Distance between two maximums determines the separation of the components. The orientation of bands allows obtaining of the position angle. The amplitude is associated with a difference of brightness between the components.

We use EMCCD (electron-multiplying CCD) to take images at the BTA speckle interferometer. Thus, the image of faint object represents a set of separate points where the quanta of light fall. The lower the magnitude of the object, the bigger number of images is required to obtain the final image.

As a matter of fact, objects with magnitude up to $15^m$ have been studied by the EMCCD-based Speckle Interferometer at the BTA telescope [Maksimov et al, 2009].

An advantage of the speckle interferometry method in comparison with the adaptive optics is its application not only to the infrared images, but to the visible region and near ultraviolet [Maksimov et al, 2009]. Moreover, the speckle interferometry equipment is significantly cheaper than adaptive optics for a large telescope, while the atmospheric seeing conditions requirements for the speckle interferometry observations are less demanding than for the observations with the use of adaptive optics.

The speckle interferometry method is described in the papers by [Labeyrie, 1970; Balega et al, 2002]. The technical description of the BTA speckle interferometer is given in [Maksimov et al, 2009].

**Binary asteroid 22 Kalliope speckle interferometry**

During the time period from December 10, 2011 to December 16, 2011, we observed the 22 Kalliope main belt binary asteroid using the speckle interferometer of the BTA operated by SAO RAS (http://www.sao.ru/Doc-k8/Telescopes/bta/descrip.html). As a result, we obtained 9 positions of the satellite Linus orbiting its main component 22 Kalliope. One of the reconstructed images is presented in Fig. 1b. It is necessary to note that relative positions were obtained from power spectrum.

**Table 1.** The Linus observational data: separation, positional angle, magnitude difference, their errors.

| Date, time | $\rho$, mas | $\sigma_\rho$, mas | $\theta$, deg | $\sigma_\theta$, deg | $\Delta m$, mag | $\sigma_{\Delta m}$, mag | $\lambda$, nm |
|---|---|---|---|---|---|---|---|
| 2011 December 10, $21^h\ 59^m$ | 398 | 16 | 336.2 | 2.3 | 3.21 | 0.17 | 800 |
| 2011 December 11, $01^h\ 14^m$ | 567 | 10 | 345.9 | 1.0 | 3.20 | 0.11 | 800 |
| 2011 December 12, $23^h\ 42^m$ | 724 | 22 | 171.1 | 1.7 | 3.18 | 0.14 | 800 |
| 2011 December 13, $02^h\ 36^m$ | 801 | 13 | 174.0 | 1.0 | 3.34 | 0.14 | 800 |
| 2011 December 13, $21^h\ 26^m$ | 530 | 7 | 195.2 | 0.8 | 3.35 | 0.08 | 800 |
| 2011 December 15, $20^h\ 52^m$ | 509 | 4 | 33.5 | 0.8 | 3.05 | 0.09 | 800 |
| 2011 December 16, $00^h\ 49^m$ | 181 | 7 | 86.0 | 2.2 | 2.47 | 0.13 | 600 |
| 2011 December 16, $00^h\ 54^m$ | 184 | 7 | 86.9 | 2.1 | 3.03 | 0.13 | 800 |
| 2011 December 16, $03^h\ 27^m$ | 236 | 7 | 128.9 | 1.7 | 3.19 | 0.14 | 800 |

Table 1 contains relative positions of the Linus as well as the apparent magnitudes difference of the satellite and the main component. The observations were made at the wavelengths of 800 nm and 600 nm. Half-width of the filter for the wavelength of 800 nm makes up 100 nm, and 40 nm for the wavelength of 600 nm correspondingly.

**Determination of Linus' apparent orbital parameters**

The observations of 22 Kalliope were carried out during several days. For this time period, the equatorial coordinates of Kalliope have changed: the right ascension has decreased on 1°.3 and the declination has increased on 0°.5. The motion of Kalliope led to change of inclination of Linus orbit plane to the picture plane. Also, the zero-point of positional angle (local direction to celestial pole) changed. The geocentric distance of the Kallope – Linus system has decreased on 0.012 a.u. For the apparent ellipse construction, the observed relative coordinates of Linus were reduced to the mean time moment of the observations $T_{\text{mean}}$ (2011 December 14, $00^{\text{h}} 04^{\text{m}}$) and corresponding astrocentric coordinate system with origin at $\alpha = 6^{\text{h}} 18^{\text{m}} 18\overset{s}{.}73$, $\delta = 32° 02' 04''\!.5$. For this, four corrections were applied to the observed positions:

- correction for phase effect,
- correction for geocentric distance change (scale),
- correction to positional angle for change of local direction to celestial pole,
- correction for change of inclination of Linus radius-vector to the picture plane.

The last correction was calculated using preliminary Linus orbit constructed without any corrections. For the correction to positional angle, the ephemerides of 22 Kalliope asteroid were calculated using EPOS software package [L'vov, Tsekmeister, 2012] with the asteroid orbital elements shared by E.Bowell (ftp://ftp.lowell.edu/pub/elgb/astorb.html). The corrections for the phase angle were calculated using Lambert's cosine law for spherical body [Toulmonde et al, 1994]. We used Kalliope's size of 166.2 km from [Descamps et al, 2008]. The phase angle of Kalliope was small (about 6°) during the observational period. Consequently, the corrections for phase were small too: not more than 0.9 mas for *X* and 3.5 mas for *Y*. We have neglected correction for non-spherical form of Kalliope because the value of this effect is very small in comparison with other effects.

Absolute values of summary corrections to *X* and *Y* Linus coordinates were from 0.5 to 6 mas.

Using the 9 positions of Linus that we observed in reference to Kalliope, we were able to plot an apparent ellipse that is a projection of the Linus orbit to the tangent plane. We used a least squares method with regard to weights in order to calculate the ellipse parameters. The weights were assigned to each position taking into account observational errors (see Table 1). In order to measure the weights, we calculated the corresponding errors for *x* and *y* coordinates in the coordinate system *Oxy* of the apparent ellipse. The centre of this coordinate system is the centre of the apparent ellipse, and the *x*-axis is directed as the semi-major axis.

We used the following formula to measure the weights:

$$p_i = \frac{1}{\sigma_{xi}^2 + \sigma_{yi}^2}, \quad i = 1, ..., 9.$$

The initial Linus positions are given in an astrocentric coordinate system *TXY* (see Fig. 2a) which centre is the photocentre of the Kalliope image. However, the centre of the apparent Linus orbit might be shifted in regard to the major component photocentre.

Thus, the ellipse that we had to determine was given in the coordinate system *Oxy* that had been shifted in regard to the astrocentric coordinate system *TXY* by unknown value $X_c$, $Y_c$ and rotated in regard to it by an unknown angle φ.

Semi-axes *a* and *b* of the apparent ellipse were also unknown. We consecutively varied all 5 unknown parameters with a predefined step within the range of possible errors. We calculated *x* and *y* for current values of parameters in *Oxy* system at each step:

$$x' = (x - X_c)\cos\varphi - (y - Y_c)\sin\varphi,$$
$$y' = -(x - X_c)\sin\varphi + (y - Y_c)\cos\varphi,$$
$$x = [-a, ..., a], \quad y = \pm b\sqrt{1 - \frac{x^2}{a^2}}.$$

At each step we calculated a $\delta^2$ squared distance between the obtained ellipse and 9 observed positions. We considered an ellipse with a minimal squared distance to be the one that we had to determine. The resultant error of fitting the ellipse into the initial position was calculated the following way:

$$\sigma = \sqrt{\frac{\delta^2 N}{\sum p_i (N-1)}},$$

As a result we plotted the ellipse that was in perfect agreement with the initial data with the error of σ = 5 mas and parameters that are given in Table 2. Fig. 2b illustrates the plotted ellipse and the positions observed.

**Table 2.** Parameters of apparent ellipse and their standard deviations

| *a*, mas | *b*, mas | $X_c$, mas | $Y_c$, mas | φ, grad |
|---|---|---|---|---|
| [$\sigma_{1a}$] | [$\sigma_{1b}$] | [$\sigma_{1x}$] | [$\sigma_{1y}$] | [$\sigma_{1\varphi}$] |
| 907.2 | 176.7 | −2.9 | 0.5 | 0.5 |
| 1.9 | 0.5 | 0.3 | 0.9 | 0.1 |

For the correct estimation of the parameter errors, we needed a number of independent determinations in order to study the distributions of the obtained parameters and their correlations. But in our case, the ellipse parameters were calculated using only nine positions while the minimum number of points to determine the ellipse is five. Thus, our observational material does not allow us to perform two independent determinations. The parameters of the ellipses that were determined on subsets of the available points showed small dispersion of parameters (see Table 2.), but the determinations were not independent and it is impossible to study the correlations. Therefore, we consider these formal errors as questionable and we accept the error of σ = 5 mas as the accuracy of the ellipse parameters *a* and *b* (see Table 3).

**Direct geometric method for true ellipse determination according to its projection**

Thus, by using Linus' relative positions, we were able to plot the apparent ellipse orbited by the satellite around its main component 22 Kalliope. We defined its semi-major and semi-minor axes, the major axis orientation in regard to the North Polar local direction (the *TY* axis of the astrocentric coordinate system, see Fig. 2b) and the ellipse centre position.

In order to define the parameters and orientation of the true ellipse we applied the direct geometric method, developed by Kisselev A.A. in 1990 year that successfully was used for a calculation visual binary stars orbits [ Kisselev A.A., 1990; Kisselev A.A., 1997]. To determine binary asteroid orbit we used this method for the first time.

The true and the apparent orbits of the binary system are the flat sections of some elliptic cylinder which axis is directed to the binary system. The apparent ellipse that lies on the tangent plane is the perpendicular cylinder section, while the true ellipse is the inclined cylinder section. The system's main body located in the true ellipse focal and its projection on the tangent plane lie on the same direct line that is parallel to the cylinder axis.

Our task here is to measure the true ellipse parameters and orientation. The solution is based on the canonical equation of the orbit apparent ellipse (1). The ellipse parameters are measured reliably if the observation covers no less than a half of the rotation period.

Let us assume that the centres of the apparent and true orbits are in the *O* point which is considered as an origin of coordinates of two coordinate systems: in the apparent ellipse plane (*Oxyz*) and in the true ellipse plane (*O*ξηζ). The *Ox* and *O*ξ axes are directed the same way as the semi-major axes of the corresponding ellipses. Thus, the equations of the apparent and true ellipses look the following way in corresponding coordinate systems:

$$b^2 x^2 + a^2 y^2 = a^2 b^2; z = 0$$
$$B^2 \xi^2 + A^2 \eta^2 = A^2 B^2; \zeta = 0 \tag{1}$$

*a, b, A, B* at this point are the semi-axes of the ellipses given in arc seconds (as required by the problem statement). Transformation of the *xyz* coordinates to ξηζ is defined by the Euler angles *i*, *w* and ω (see Fig. 3) where:

*i* is the angle between the *O*ζ and *Oz* axes, and defines the true ellipse inclination to the tangent plane;

*w* is the angle between the *Ox* axis and the line of nodes *O*Ω (an intersection line of the orbital plane and the tangent plane), and defines the true ellipse orientation in the tangent plane;

ω is the angle between the *O*ξ axis and the line of nodes *O*Ω, and defines the true ellipse orientation in its own plane.

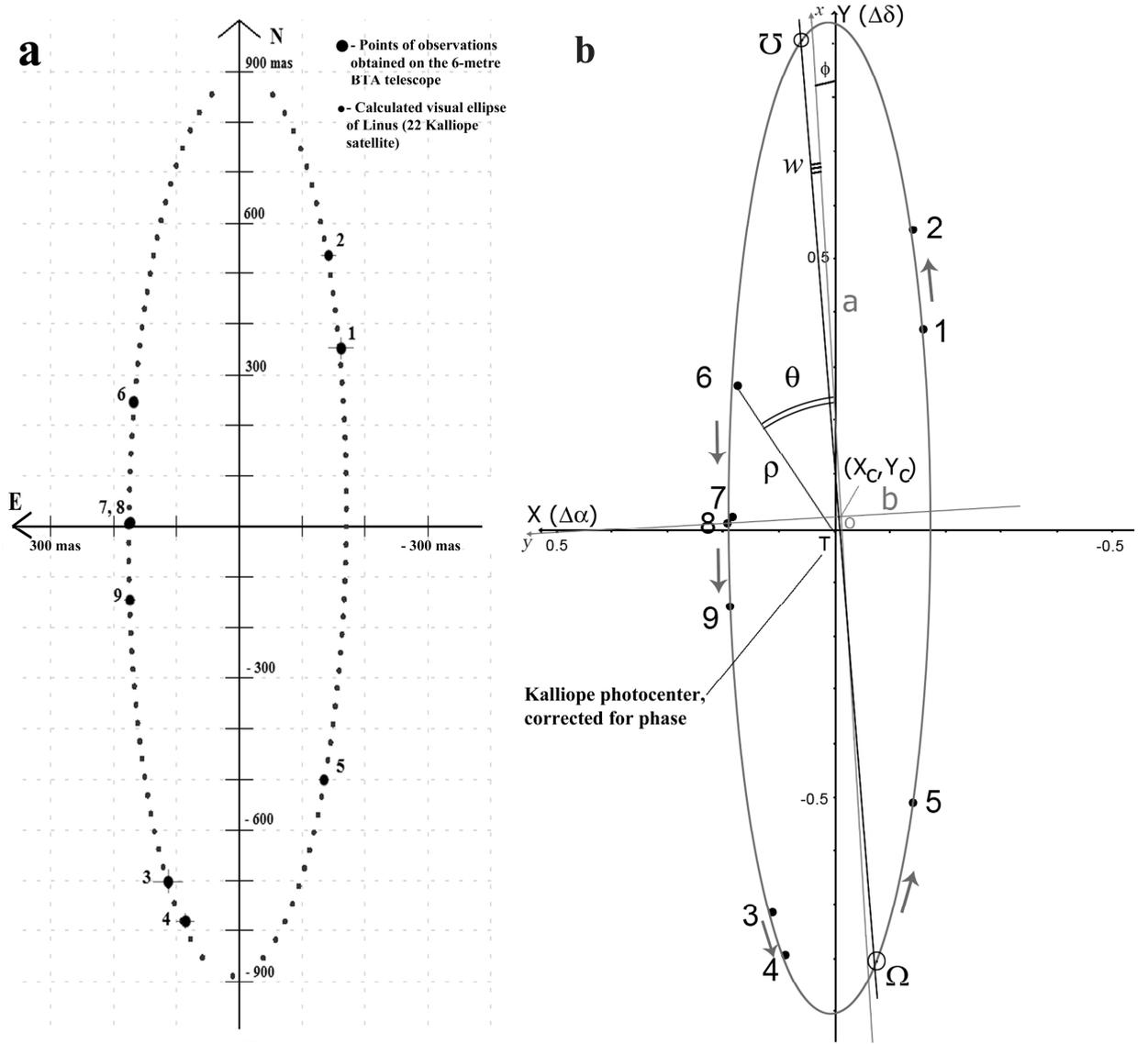

**Fig. 2. a)** Observed Linus positions with error bars and the calculated apparent ellipse. **b)** Two coordinate systems: astrocentric system *TXY* and apparent ellipse coordinate system *Oxy*.

The corresponding coordinate transformations can be presented through the rotation matrix $R(w, i, \omega)$ obtained as a result of three consecutive rotations of coordinate trihedral axes: a rotation about the *Oz* axis through a $(-w)$ angle, a rotation about the *Ox* axis through an *i* angle, and a rotation about the new *Oz* axis position through a $+\omega$ angle.

$$r_{\xi\eta\zeta} = Z(\omega) X(i) Z(-w) r_{xyz} \qquad (2)$$

In this case we use the following rotation matrices:

$$Z(-w) = \begin{matrix} \cos w & -\sin w & 0 \\ \sin w & \cos w & 0 \\ 0 & 0 & 1 \end{matrix} \qquad X(i) = \begin{matrix} 1 & 0 & 0 \\ 0 & \cos i & \sin i \\ 0 & -\sin i & \cos i \end{matrix} \qquad Z(\omega) = \begin{matrix} \cos \omega & \sin \omega & 0 \\ -\sin \omega & \cos \omega & 0 \\ 0 & 0 & 1 \end{matrix}$$

$r_{\xi\eta\zeta}$ and $r_{xyz}$ are the vectors that consist of coordinates $\xi$, $\eta$, $\zeta$ and $x, y, z$.

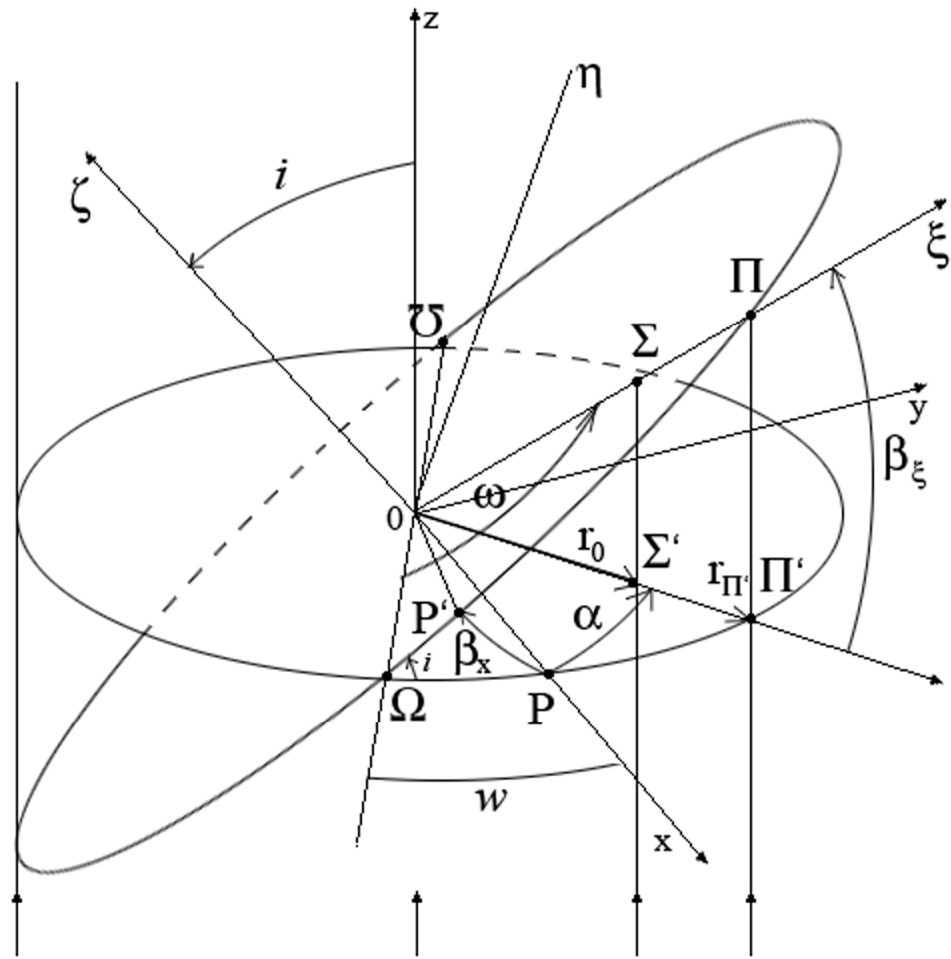

**Fig. 3.** The direct geometric method illustration. Mutual orientation of apparent and true ellipses.

We use the (2) transformation to record the cylinder equation in the true ellipse coordinate system ($O\xi\eta\zeta$). The transformation consists of two stages. At first, we switch from the $Oxyz$ system to the $Ox'y'z'$ system that is rotated through the ($-w$) angle with regard to the initial one, in other words we perform the transformation:

$$x = x'\cos w + y'\sin w$$
$$y = -x'\sin w + y'\cos w$$
$$z = z'$$

We get:
$$a_{11}x'^2 + 2a_{12}x'y' + a_{22}y'^2 = a^2 b^2$$

At this point:

$$a_{11} = a^2 \sin^2 w + b^2 \cos^2 w$$
$$a_{12} = -\sin w \cos w(a^2 - b^2)$$
$$a_{22} = a^2 \cos^2 w + b^2 \sin^2 w$$

At the second stage we switch from the $Ox'y'z'$ system to the $O\xi\eta\zeta$ system i.e. we perform the following transformation:
$$x' = \xi \cos\omega - \eta \sin\omega$$
$$y' = \xi \sin\omega \cos i + \eta \cos\omega \sin i - \zeta \sin i$$
$$z' = \xi \sin\omega \sin i + \eta \cos\omega \sin i + \zeta \cos i$$

We get:
$$A_{11}\xi^2 + 2A_{12}\xi\eta + A_{22}\eta^2 + 2A_{13}\xi\zeta + 2A_{23}\eta\zeta + A_{33}\zeta^2 = a^2 b^2 \qquad (3)$$

In cylinder equations (3) and (4) we set $\zeta = 0$, thus we calculate the cylinder section that goes through the coordinate origin and is inclined by a $\pm i$ angle with regard to the $Oxy$ plane. We record this section equation as follows:
$$A_{11}\xi^2 + 2A_{12}\xi\eta + A_{22}\eta^2 = a^2 b^2 \qquad (4)$$

$A_{jk}$ at this point are the true ellipse parameters that depend on the apparent ellipse parameters and orientation angles $i$, $w$, $\omega$. We record these dependencies in the explicit way:
$$A_{11} = a_{11}\cos^2\omega + 2a_{12}\sin\omega\cos\omega\cos i + a_{22}\sin^2\omega\cos^2 i$$
$$A_{12} = -a_{11}\sin\omega\cos\omega + a_{12}(\cos^2\omega - \sin^2\omega)\cos i + a_{22}\sin\omega\cos\omega\cos^2 i \qquad (5)$$
$$A_{22} = a_{11}\sin^2\omega - 2a_{12}\sin\omega\cos\omega\cos i + a_{22}\cos^2\omega\cos^2 i$$

We use the expressions (5) to derive the formulae that bind the apparent and true ellipses parameters (1) together with the $i$, $w$, $\omega$ angles.

According to the condition for the $O\xi$ axis that by construction has to be in line with the semi-major axis of the true ellipse (see formulae (1) and (4); note that in formula (4) factor $A_{12}$ has to equal 0):
$$A_{12} = 0$$

That leads to the following condition for $\omega$:
$$\mathrm{tg}2\omega = \frac{2a_{12}\cos i}{a_{11} - a_{22}\cos^2 i} \qquad (6)$$

From (6) we easily get the expressions for $\sin 2\omega$, $\cos 2\omega$, $\sin^2\omega$, $\cos^2\omega$:
$$\sin 2\omega = \frac{2a_{12}\cos i}{a^2\kappa}$$
$$\cos 2\omega = \frac{a_{11} - a_{12}\cos^2 i}{a^2\kappa}$$
$$\sin^2\omega = \frac{-(a_{11} - a_{22}\cos^2 i) + a^2\kappa}{2a^2\kappa} \qquad (7)$$
$$\cos^2\omega = \frac{(a_{11} - a_{22}\cos^2 i) + a^2\kappa}{2a^2\kappa}$$
$$a^2\kappa = -\sqrt{(a_{11} - a_{22}\cos^2 i)^2 + 2a_{12}\cos^2 i}$$

We keep the «–» sign in front of the square root in order to provide the same reference direction for the $w$ and $\omega$ angles from the line of nodes to the $Ox$ and $O\xi$ axes. We use the obtained expressions to transform the expressions for $A_{11}$ and $A_{22}$:

$$A_{11} = \frac{1}{2}\left[(a_{11} + a_{22}\cos^2 i) + a^2\kappa\right] = \frac{a^2}{2}(f+\kappa)$$

$$A_{22} = \frac{1}{2}\left[(a_{11} + a_{22}\cos^2 i) - a^2\kappa\right] = \frac{a^2}{2}(f-\kappa)$$

At this point we introduce the following notation: $a^2 f = a_{11} + a_{22}\cos^2 i$.

We compare the coefficients in equations (4) and (1) that have to determine the same true ellipse in the $O\xi\eta$ system. We set the following identities that are essential for this task:

$$A^2 = \frac{2b^2}{f+\kappa} \qquad f = b^2\frac{A^2+B^2}{A^2 B^2}$$
$$B^2 = \frac{2b^2}{f-\kappa} \qquad ; \qquad \kappa = -b^2\frac{A^2-B^2}{A^2 B^2} \qquad (8)$$

From the expression for $a^2\kappa$ we get:

$$f^2 - \kappa^2 = 4\frac{b^2}{a^2}\cos^2 i \qquad (9)$$

and, therefore

$$\sin^2 i = \frac{A^2 B^2 - a^2 b^2}{A^2 B^2}$$
$$\cos^2 i = \frac{a^2 b^2}{A^2 B^2} \qquad (10)$$

The formula (10) can be used to control the accuracy of the performed derivations as it expresses a geometric property of the squares ratio of two elliptic cylinder sections; one of this sections is perpendicular to the axis.

We derive one more intermediary formula that is required to define the $w$ orientation angle. By transforming the expression (9) we get:

$$f = 1 + \cos^2 i - e^2 \cos^2 \beta_x \qquad (11)$$

$\beta_x$ at this point is the angle between the $Ox$ axis and the $O\xi\eta$ plane, $e$ is the eccentricity of the apparent ellipse. The following formula is true for $\sin\beta_x$:

$$\sin\beta_x = \sin w \sin i \qquad (12)$$

By combining the formulae (8), (10) – (12) we get:

$$\sin^2 \beta_x = \frac{b^2(a^2-B^2)(A^2-a^2)}{(a^2-b^2)A^2 B^2}$$
$$\sin^2 w = \frac{b^2}{(a^2-b^2)}\frac{(a^2-B^2)(A^2-a^2)}{(A^2 B^2 - a^2 b^2)} \qquad (13)$$

We calculate the trigonometric functions of the $\omega$ and $\beta_\xi$ angles in the same way. These functions are bound by the expression:

$$\sin\beta_\xi = \sin\omega \sin i \qquad (14)$$

In order to calculate $\sin\omega$ we use the formulae (7), (8), (11); in order to calculate $\sin\beta_x$ we use the expression (13). We get:

$$\sin^2 \beta_\xi = \frac{(A^2 - b^2)(A^2 - a^2)}{A^2(A^2 - B^2)}$$

$$\cos^2 \beta_\xi = \frac{A^2(a^2 - B^2) + b^2(A^2 - a^2)}{A^2(A^2 - B^2)} \qquad (15)$$

$$\sin^2 \omega = \frac{B^2}{(A^2 - B^2)} \frac{(A^2 - b^2)(A^2 - a^2)}{(A^2 B^2 - a^2 b^2)}$$

The formulae (10), (13) and (15) solve the problem of defining the orientation of the inclined elliptic cylinder section, in case we know its main parameters — the semi-axes $A$ and $B$. In order to measure these parameters we use the distance $r_0$ from the apparent ellipse centre to the main body projection (the segment $O\Sigma'$ in Fig. 3), and $\alpha$ which is the angle between $r_0$ and the $Ox$ axis in tangent plane.

If we continue the radius $r_0$ up to the intersection with the apparent ellipse arc, we will be able to measure the $\Pi'$ point that is the projection to the tangent plane of the true orbit periastron $\Pi$. We indicate the radius of the $\Pi'$ point as $r_{\Pi'}$; it is easy to calculate its value as we know the $\alpha$ angle:

$$r_{\Pi'} = \frac{b^2}{1 - e^2 \cos^2 \alpha} = \frac{a^2 b^2}{a^2 \sin^2 \alpha + b^2 \cos^2 \alpha}$$

According to the projection condition:

$$r_{\Pi'} = A^2 \cos^2 \beta_\xi \qquad r_0 = A^2 E^2 \cos^2 \beta_\xi \qquad (16)$$

where $E$ is the true ellipse eccentricity, and the $\beta_\xi$ angle is defined in accordance with (14). We use the condition (16) to calculate the true ellipse eccentricity:

$$E^2 = \frac{r_0^2}{r_{\Pi'}^2} = \frac{r_0^2}{b^2}(1 - e^2 \cos^2 \alpha) \qquad (17)$$

Equation to calculate the true ellipse semi-major axis is derived from the formula (16) for $r_0$ where we use the expressions from (15) for $\cos \beta_\xi$:

$$A^4(1 - E^2) - A^2(a^2 + b^2 - r_0^2) + a^2 b^2 = 0 \qquad (18)$$

The discriminant of the obtained biquadratic equation is the following:

$$D = (a^2 + b^2 - r_0^2)^2 - 4(1 - E^2)a^2 b^2 = a^4 + b^4 + r_0^4 - 2a^2 b^2 - 2a^2 r_0^2 - 2b^2 r_0^2 + 4E^2 a^2 b^2.$$

The last summand in the expression for the discriminant is estimated as:

$$4E^2 a^2 b^2 > 4 r_0^2 b^2 \quad \text{because} \quad E^2 = \frac{r_0^2}{r_{\Pi'}^2} > \frac{r_0^2}{a^2}$$

Therefore, we can set the continued inequalities:

$$(a^2 + b^2 - r_0^2)^2 > D > (a^2 - b^2 - r_0^2)^2 > 0$$

That leads to the fact that a biquadratic equation has two real positive roots. The solution is the one that satisfies the projection condition $A > a$.

According to the derived algorithm the inclined true ellipse parameters are defined rigorously in geometric sense. Ambiguity when defining the $w$ and $\omega$ angles is controlled by the formula $\cos(w + \alpha)\cos \beta_\xi = \cos \omega$, where the angle $\alpha$ is predefined by the value and the reference direction.

Non-correctable ambiguity is persistent only when defining the inclination $i$ ($i$ or 180–$i$), due to the fact that two different ellipses symmetrically inclined to tangent plane have the same projection, and it's necessary to choose one of the options, to decide which node is ascending (i.e. to decide in which node the satellite will move away from the observer). According to the choice of the ascending node we need to add 180° to the ω angle.

### Linus orbit calculation

Initial data for measuring true ellipse parameters and orientation can be found in Table 3. At first, we calculate the eccentricity $E$ using the formula (17), after that we calculate the semi-major axis $A$ according to the equation (18). Inclination $i$ is obtained from the expression (10). We calculate the periastron longitude ω using formula (7) for $\sin^2\omega$, and the line of nodes position angle $w$ according to the formula (13) for $\sin^2 w$.

Table 3. Initial data for direct geometrical method

| | | |
|---|---|---|
| Semi-major axis $a$ | mas | 907.2 ± 5 |
| Semi-minor axis $b$ | mas | 176.7 ± 5 |
| Positional angle of semi-major axis $a$ | deg | 0.5 ± 0.4 |
| Distance between apparent ellipse centre and photocenter of 22 Kalliope | mas | 2.5 ÷ 3.4 |
| α — angle between the of semi-major axis $a$ and direction toward photocenter of 22 Kalliope | deg | 60 ÷ 100 |

It is important to note when measuring orbit orientation that $i$ is the inclination of the true orbit to the tangent plane which is defined by its cosine: $\cos i = \vec{Z}\cdot\vec{Q} = Q_3$. $Z$ at this point is the unit normal vector that is perpendicular to the tangent plane and is in the same direction as the observer.

Inclination is defined unequivocally: 0° < $i$ < 90°, if $\cos i > 0$ then the direction of the orbital apparent satellite motion has the same direction with the rotation from $X$ to $Y$ axis in tangent plane, $\dot\theta < 0$; 90° < $i$ < 180°, if $\cos i < 0$ then satellite orbital motion has the opposite direction, $\dot\theta > 0$.

Table 4. The true ellipse parameters

| | | |
|---|---|---|
| Semi-major axis $A$ | mas | 907.3 ± 5 |
| Semi-minor axis $B$ | mas | 907.2 ± 5 |
| Eccentricity $E$ | | 0.016 ± 0.004 |
| Inclination $i$ | deg | 78.77 ± 0.1 |
| $w$ — angle between nodes line and $Ox$ axis | deg | 0.5 ± 0.1 |
| Ω — ascending node positional angle ($w$ + 180°) | deg | 180.5 ± 0.1 |
| Periastron longitude ω (from Ω) | deg | 88.14 ± 3 |

We calculated the periastron passage time $T_P$ the following way: relative coordinates of the observed position were transformed to the apparent ellipse coordinate system. The node line positional angle was added to the position angle of the apparent ellipse semi-major axis $w = 0°.5 + 0°.0001 = 0°.5001$.

Next, in order to transform the coordinates we rotated the coordinate system along the $z$ axis through $-0.5°$ and the $Y'$ axis got in line with the line of nodes. Then we performed a rotation about the $Y'$ axis through the $i$ angle, and calculated the Linus position angles in the true orbit coordinate system. In order to calculate the periastron passage time we composed the following expressions for the position angles $\theta$ and corresponding moments $t$: $T_P = t + (\omega - \theta)/n$. After that we calculated the mean for $T_P$ taking the phase into account. Thus, we obtained the periastron passage time equal to $T_P = 2011.943593 \pm 0.000016$ years. Adopted value of $n$ is derived from period of $3.595712 \pm 6.8 \cdot 10^{-5}$ days [Vachier et al, 2012] and equal to 36567.80632 °/yr.

From the available number of Linus positions revolving around 22 Kalliope obtained in the period from December 10, 2011 to December 16, 2011 on the 6-m BTA telescope we were unable to determine the period due to large ambiguity of the resulting estimates. Therefore, to calculate the Linus orbit we used period P = $3.595712 \pm 6.8 \cdot 10^{-5}$ days from [Vachier et al, 2012].

The true ellipse orientation parameters were obtained in the local astrocentric coordinate system corresponding to the mean epoch of observation (see Table 4); they were transformed into ecliptic coordinate system.

It is necessary to note that while calculating Linus' orbit we also took into account the corrections for light velocity and asteroid displacement in regard to the Earth during the observation period. These corrections did not exceed 3 mas.

Hence, we obtained the orbital parameters for Linus. In Table 5 you can see our findings compared with the results obtained by other researchers [Descamps et al, 2008; Vachier et al, 2012]. It is obvious that the results of this research are in agreement with the previous findings.

It is important to note that the eccentricity of 0.016 that we obtained in this research is within the error range of 0.03, and thus is in agreement with the value presented in the paper by Vachier [Vachier et al, 2012]. The difference of orbit pole ecliptic latitude $\beta$ from the value obtained by Descamps [Descamps et al, 2008] may be caused by some orbit plane precession.

Table 5. Linus orbital parameters in ecliptic coordinate system

| Orbital parametres | Our results | P.Descamps et al, 2008 | F. Vachier et al, 2012 |
|---|---|---|---|
| Semi-major axis | $a = 907.3 \pm 5$ mas = $1109 \pm 6$ km | $1099 \pm 11$ km | $1082 \pm 33$ km |
| Semi-minor axis | $b = 907.2 \pm 5$ mas = $1109 \pm 6$ km | | |
| Eccentricity | $e = 0.016 \pm 0.004$ | | $0.0069 \pm 0.03$ |
| Periastron longitude (in local astrocentric system) | $\omega = 88° \pm 3°$ | | |
| Periastron passage time | | | |
|   | $T_P$ = 2011.94359 ± 0.00002 year |   |   |
| Ecliptic coordinate of the orbital pole | β = −11° ± 2°<br>λ = 190° ± 2° | β = −3° ± 2°<br>λ = 197° ± 2° |   |
| Ecliptic coordinate of the periastron | β = −51° ± 3°<br>λ = 295°± 3° |   |   |
| Longitude of the ascending node | λ = 281° ± 1° | 284.5° ± 2.0° | 285.05° ± 2.11° |
| Inclination | $i$ = 101° ± 1° | 99.4° ± 0.5° | 94.18° ± 1.92° |

## Conclusion

Thus, we obtained instantaneous orbit of the 22 Kalliope satellite for the mean epoch of observation (2011 December 14, 00h 04m) using the speckle interferometry method for separating asteroid components and the direct geometric method for calculating of the true ellipse of an orbit. As a result, we have obtained the full range of the elements of Linus orbit; furthermore, the orbital elements that were previously measured by other researches are in good agreement with the results of this research. However, slight differences might attest to the fact that Linus' orbit experiences moderate perturbations. The suggested method of instant orbital calculation of asteroid satellites will allow monitoring changes in the measured orbits over the course of time.

The present work is the first stage of study of the binary asteroid 22 Kalliope based on the observed data from the 6-m BTA telescope with the application of the speckle interferometry method. In the future we are planning to take into account the shape of the main component 22 Kalliope. We are going to continue speckle observations of this binary asteroid with aim to improve Linus orbital parameters and to study their changes.

## Acknowledgements

This research was funded by:

The Ministry of Education and Science of the Russian Federation under the programme "Research and Pedagogical Specialists of Innovative Russia 2009-2013" (government grant 8704);

The Russian Federation President Grant for Government Support of Young Russian Scientists — Candidates of Sciences (MK-1001.2012.2);

The Russian Foundation for Basic Research (project no. 12-02-00675a).

## References:

Balega, I. I., Balega, Y. Y., Hofmann, K.-H., Maksimov, A. F., Pluzhnik, E. A., Schertl, D., Shkhagosheva, Z. U., Weigelt, G., 2002. Speckle interferometry of nearby multiple stars. Astron. Astrophys. 385, 87-93.

Balega Y. Y., Dyachenko V. V., Maksimov A. F., Malogolovets E. V., Rastegaev D. A., Romanyuk I. I., 2011. Speckle interferometry of magnetic Ap/Bp stars at the BTA 6-m telescope (new binary and multiple systems). Astronomische Nachrichten, Vol.332, Issue 9/10, p.978-982

Cellino, A., Diolaiti, E., Ragazzoni, R., Hestroffer, D., Tanga, P., Ghedina, A., 2003. Speckle interferometry observations of asteroids at TNG. Icarus 162 (2), 278-284.

Cellino, A., 2005. Ground-Based Optical Observations of Asteroids. Highlights of Astronomy, Vol. 13, as presented at the XXVth General Assembly of the IAU - 2003 [Sydney, Australia, 13 - 26 July 2003]. Edited by O. Engvold. San Francisco, CA: Astronomical Society of the Pacific, ISBN 1-58381-189-3. XXIX + 1085, 752.

Descamps, P., Marchis, F., Pollock, J., Berthier, J., Vachier, F., Birlan, M., Kaasalainen, M., Harris, A. W., Wong, M. H., Romanishin, W. J., Cooper, E. M., Kettner, K. A., Wiggins, P., Kryszczynska, A., Polinska, M., Coliac, J.-F., Devyatkin, A., Verestchagina, I., Gorshanov, D., 2008. New determination of the size and bulk density of the binary Asteroid 22 Kalliope from observations of mutual eclipses. Icarus 196 (2), 578-600.

Descamps, P., 2005. Orbit of an astrometric binary system. Celestial Mechanics and Dynamical Astronomy, 92, 381-402.

Grosheva, E. A., 2006. Analysis of periodic perturbations in the multiple system ADS 15571. Astrophysics 49 (3), 397-404.

Kaasalainen M., Torrpa J., Piironen J., 2002. Models of twenty asteroids from photometric data. Icarus 159, 369-395.

Kisselev, A. A., 1990. The determination of elliptical orbit of binary star by its projection on the celestial sphere, Astronomical and Geodesic research: nearby binary and multiple stars. The collection of scientific Papers, Sverdlovsk, 36.

Kisselev, A. A. A Direct Geometrical Method for Determination of the Elliptic Orbit of a Binary Star Using its Projection on the Celestial Sphere. In Visual Double Stars : Formation, Dynamics and Evolutionary Tracks. Edited by J.A. Docobo, A. Elipe, and H. McAlister. Dordrecht : Kluwer Academic, 1997., 357.

Labeyrie, A., 1970. Attainment of Diffraction Limited Resolution in Large Telescopes by Fourier Analysing Speckle Patterns in Star Images. Astron. Astrophys. 6, 85.

L'vov, V. N.; Tsekmeister, S. D., 2012. The use of the EPOS software package for research of the solar system objects, Solar System Research, Volume 46, Issue 2, pp.177-179.

Maksimov, A. F., Balega, Yu. Yu., Dyachenko, V. V., Malogolovets, E. V., Rastegaev, D. A., Semernikov, E. A., 2009. The EMCCD-based speckle interferometer of the BTA 6-m telescope: Description and first results. Astrophysical Bulletin 64 (3), 296-307.

Marchis, F., Descamps, P., Hestroffer, D., Berthier, J., Vachier, F., Boccaletti, A., de Pater, I., Gavel, D., 2003. A three-dimensional solution for the orbit of the asteroidal satellite of 22 Kalliope. Icarus, 165 (1), 112-120.

Marchis, F., Descamps, P., Hestroffer, D., Berthier, J., de Pater, I., 2004. Fine Analysis of 121 Hermione, (45) Eugenia, and (90) Antiope Binary Asteroid Systems With AO Observations. American Astronomical Society, DPS meeting #36, #46.02; Bulletin of the American Astronomical Society, 36, 1180.

Marchis F., Descamps P., Hestroffer D., Berthier J., 2005. Discovery of the triple asteroidal system (87) Sylvia. Nature 436, 822-824.

Marchis F., Kaasalainen M., Hom E., 2006. Shape, size and multiplicity of main-belt asteroids. Icarus, 185 (1), 39-63.

Marchis, F., Descamps, P., Baek, M., Harris, A. W., Kaasalainen, M., Berthier, J., Hestroffer, D., Vachier, F., 2008. Main belt binary asteroidal systems with circular mutual orbits. Icarus 196 (1), 97-118.

Masiero J. R., Mainzer, A. K., Grav, Bauer T., J. M., Cutri R. M., Dailey J., Eisenhardt P. R. M., McMillan R. S., Spahr T. B., Skrutskie M. F., Tholen D., Walker R. G., Wright E. L., DeBaun E., Elsbury D., Gautier T. IV, Gomillion S., and Wilkins A., et al 2011. Main Belt Asteroids with WISE/NEOWISE. I. Preliminary Albedos and Diameters. ApJ, 741, 68.

Merlin W.J., Close L.M., Dumas C., Shelton J.C., 2000. Discovery of Companions to Asteroids (762) Pulcova and (90) Antiope by Direct Imaging. American Astronomical Society, DPS Meeting #32, #13.06; Bulletin of the American Astronomical Society, 32, 1017.

Morbidelli A., Levison H., Bottke W., 2006. Formation of the binary near-Earth object 1996 FG3: Can binary NEOs be the source of short-CRE meteorites? Icarus, vol. 41, № 6, 2006, p.875-887.

Pluzhnik E. A., 2005. Differential photometry of speckle-interferometric binary and multiple stars, A&A 431, 587–596.

Prokof'eva, V. V., Tarashchuk, V. P., 1995. Development of binary asteroids concepts. Solar System Research 29 (2), 105 - 113.

Roberts, L. C., Jr., McAlister, H. A., Hartkopf, W. I., Franz, O. G., 1995. A Speckle Interferometric Survey for Asteroid Duplicity. Astronomical Journal 110, 2463.

Tedesco, E.F., Noah, P.V., Noah, M. and Price, S.D. 2002. The supplemental IRAS Minor Planet Survey, Astron.J. 123, 1056-1085.

Toulmonde, M., Chollet, E., 1994. Correction de phase et positions planetaires. Astronomy and Astrophysics 287, 1014-1020.

Vachier, F., Berthier J., Marchis F., 2012. Determination of binary asteroid orbits with a genetic-based algorithm. Astron. Astrophys. 543, id. A68, pp. 9.